\documentclass[fleqn,10pt]{wlscirep}
\usepackage[utf8]{inputenc}
\usepackage[T1]{fontenc}
\title{Record statistics of bursts signals the onset of acceleration towards failure}

\author[1]{Vikt\'oria K\'ad\'ar}
\author[1,2]{Gerg\H o P\'al}
\author[1,2,*]{Ferenc Kun}

\affil[1]{Department of Theoretical Physics, Doctoral School of Physics, 
Faculty of Science and Technology, University of Debrecen, H-4010 Debrecen, P.O.Box: 5, 
Hungary}
\affil[2]{Institute of Nuclear Research (Atomki), H-4026 Debrecen, Poroszlay \'ut 6/c, Hungary}

\affil[*]{ferenc.kun@science.unideb.hu}

\begin{abstract}
Forecasting the imminent catastrophic failure has a high importance for a large variety
of systems from the collapse of engineering constructions, through the emergence of landslides and 
earthquakes, to volcanic eruptions. Failure forecast methods predict the lifetime of the system
based on the time-to-failure power law of observables describing the final acceleration towards failure.
We show that the statistics of records of the event series of breaking bursts, 
accompanying the failure process, provides a powerful tool to detect the onset of acceleration,
as an early warning of the impending catastrophe. 
We focus on the fracture of heterogeneous materials using a fiber bundle model, 
which exhibits transitions between perfectly brittle, quasi-brittle, 
and ductile behaviors as the amount of disorder is increased.
Analyzing the lifetime of record size bursts,  
we demonstrate that the acceleration starts at a characteristic record rank, 
below which record breaking slows down due to the dominance 
of disorder in fracturing, while above it stress redistribution gives rise to an 
enhanced triggering of bursts and acceleration of the dynamics. The emergence of this signal 
depends on the degree of disorder making both highly brittle fracture of low disorder
materials, and ductile fracture of strongly disordered ones, unpredictable.
\end{abstract}
\begin{document}

\flushbottom
\maketitle

\thispagestyle{empty}

\section*{Introduction}
Forecasting failure is a long standing problem which has an utmost importance to mitigate 
the consequences of the collapse of engineering constructions and of natural catastrophes like landslides,
earthquakes, volcanic eruptions, rock and snow avalanches \cite{voight_method_1988,voight_relation_1989,main_ffm_1999,GJI:GJI1884,alava_lifetime_pre2016,
pradhan_frontiers_2019,tarraga2008447,main_GJI:GJI4982}. 
Fracture processes of heterogeneous materials occurring 
under constant or slowly varying external loads play a decisive role for the emergence 
of catastrophic failures. The micro and meso-scale heterogeneity of materials has 
the consequence that their fracture process is accompanied 
by crackling noise, i.e.\ fracture proceeds in intermittent bursts of local
breakings which generate acoustic emissions \cite{dahmen_nature_2011,PhysRevLett.110.088702,salje_crackling_2014,
nataf_predicting_2014,guarino_experimental_1998,rosti_crackling_2009,alava_lifetime_pre2016}. 
Cracking bursts can be considered as precursors 
of the ultimate failure of the system, so that they can be exploited to forecast 
the impending catastrophic event 
\cite{hao_power-law_2013,vasseur_scirep_2015,pradhan_crossover_2005-1,johansen_critical_2000, uyeda_geoelectric_2000,Heap201171,
amitrano_seismic_2005,amitrano_brittle_2006,michlmayr_fiber_optic2017,
sarlis_investigation_2008,varotsos_epl_2011,sarlis_tecton_2011,nataf_predicting_2014,
sarlis_epl_2018,varotsos_identifying_2015,
landslides_book_deblasio,vives_coal_burst_2019}.
After a longer period of steady evolution, failure is approached through an acceleration 
of the dynamics which is indicated by the increasing rate of deformation, and acoustic or 
seismic signals \cite{main_GJI:GJI4982,main_ffm_1999,Heap201171,andersen_predicting_2005,sammonds_role_1992,
sammonds_bvalue_grl1998,michlmayr_fiber_optic2017,johansen_critical_2000,
nataf_predicting_2014,salje_prl_2018,petri_experimental_1994,guarino_experimental_1998,
vives_coal_burst_2019}.  
Failure forecast methods (FFM) rely on the analogy of failure and critical phenomena which implies
time-to-failure power laws of observables in the acceleration regime making possible 
to predict the lifetime of the system \cite{voight_method_1988,voight_relation_1989,hao_relation_2016,tarraga2008447,main_GJI:GJI4982,main_ffm_1999}.

Disorder is an inherent property of natural and most of the artificially made materials. 
Depending on the relevant length scale,
it appears in the form of dislocations, microcracks, flaws, grain boundaries, which 
affect the nucleation and propagation cracks.
Experimental and theoretical studies have revealed that the intensity of the precursory activity, 
and hence, the predictability of failure, 
depends on the degree of materials' disorder.  
In the limiting
case of zero disorder, the ultimate failure occurs in an abrupt way with hardly any precursors
\cite{brechet_coffin-manson_1992,zapperi_first-order_1997,sornette_scaling_1998,menezes-sobrinho_influence_2010}.
However, higher disorder makes possible to arrest propagating cracks giving rise to a 
gradual accumulation of damage with a growing rate of breaking bursts as failure is 
approached \cite{ramos_prl_2013,johansen_critical_2000,sornette_predictability_2002,saichev_andrade_2005}.
This effect has recently been precisely quantified by experiments performed 
on the compressive failure of porous glass samples where the degree of heterogeneity 
could be well controlled during the sample preparation \cite{vasseur_scirep_2015}. 
It has been demonstrated that the accuracy of the lifetime prediction of FFM rapidly improves
with increasing mesoscale heterogeneity of the material \cite{vasseur_scirep_2015}
underlining the key importance of heterogeneity for forecasting failure.

Instead of the failure time, here we focus on the onset of acceleration
which marks the start of the critical regime of the evolution of the fracture process, and hence, 
can serve as an early warning of the imminent failure, similar to the entropy change 
of seismicity under time reversal suggested earlier for earthquakes 
\cite{sarlis_epl_2018,sarlis_tecton_2011,varotsos_epl_2011,varotsos_apl_2007,sarlis_physica_2018,varotsos_natural_time_book}.
To generate fracture processes of heterogeneous materials we use a fiber bundle model
\cite{hansen2015fiber,kun_extensions_2006,hidalgo_avalanche_2009-1}, which
has the advantage that varying the amount of microscale disorder, it exhibits
transitions between distinct phases of perfectly brittle, quasi-brittle, and ductile fracture.
To quantify how the degree of disorder determines the predictability of failure, 
we investigate the internal structure of the sequence of breaking bursts analyzing
the record size events. Records are bursts of size greater than any previous crackling event 
of the fracture process so that they can easily be identified by experimental techniques both
in laboratory and in field measurements above the noisy background.
The record statistics (RS) of stochastic time series has attracted a great attention 
due to its relevance for weather, climate and earthquake research 
\cite{wergen_krug_epl_2010,davidsen_grl_2006,davidsen_pre_2008,yoder_npg_2010,miller_scaling_2013,
wergen_record_2013}.
The RS analysis has proven very successful to reveal trends, correlations, and 
spatio-temporal clustering of events in complex evolving systems \cite{wergen_krug_epl_2010,davidsen_grl_2006,
davidsen_pre_2008,yoder_npg_2010,miller_scaling_2013,scherbakov_pre_2013,wergen_record_2013,
danku_frontiers_2014,record_dem_PhysRevE.93.033006,majumdar_prl_rainfall_2019}.
For fracture processes, here we demonstrate that the waiting time between consecutive record breakings, 
i.e.\ the lifetime of records, is very sensitive 
to the details of the fracture process providing a clear signal of the acceleration 
of the dynamics towards ultimate failure. In particular,
we show the existence of a characteristic record rank $k^*$ which marks the onset of acceleration 
of record breaking: below $k^*$ record breaking slows down due to the dominance of disorder 
in the fracture process, while above it the stress redistribution gives rise to an enhanced triggering 
of bursts after breaking events. Detecting $k^*$ can be exploited as an early signal of the imminent 
ultimate failure of the system, however, the significance of the accelerating regime 
strongly depends on the degree of disorder. Most notably, we show that the highly brittle fracture 
of low disorder materials and the ductile failure of strongly disordered ones are both unpredictable 
due to the absence of accelerated record breaking. Our results imply the existence of a lower and upper 
bound of the amount of materials' disorder beyond which no meaningful failure prediction is possible.

\section*{Results}

\subsection*{Disorder driven transition between brittle, quasi-brittle, and ductile fracture}
To investigate the evolution of fracture processes leading to ultimate failure,
we use a generic fiber bundle model (FBM) which has proven successful in reproducing both the
constitutive response and the intermittent bursting dynamics of heterogeneous materials 
on the macro- and microscales, respectively \cite{danku_disorder_2016,kadar_pre_2017}. 
The model is composed of $N$ parallel fibers, similar to a rope, which have a linearly elastic behavior.
Materials' disorder is represented by
the random strength $\varepsilon_{th}^i (i=1,\ldots, N)$ of fibers with a fat-tailed probability 
distribution, i.e.\ a power law distribution is considered 
\begin{equation}
p(\varepsilon_{th})=D\varepsilon_{th}^{-(1+\mu)},
\label{eq:fstrength}
\end{equation}
over the range $\varepsilon_{min}\leq\varepsilon_{th}\leq\varepsilon_{max}$.
In the calculations the lower cutoff is fixed to $\varepsilon_{min}=1$ so that
the amount of disorder can be tuned by varying the upper cutoff $\varepsilon_{max}$
and the exponent $\mu$ of the distribution Eq.\ (\ref{eq:fstrength}),
which control the range of variability of fibers's strength and the decay of weaker against 
stronger fibers, respectively.
\begin{figure}
\begin{center}
\includegraphics[bbllx=50,bblly=30,bburx=380,bbury=330,scale=0.6]{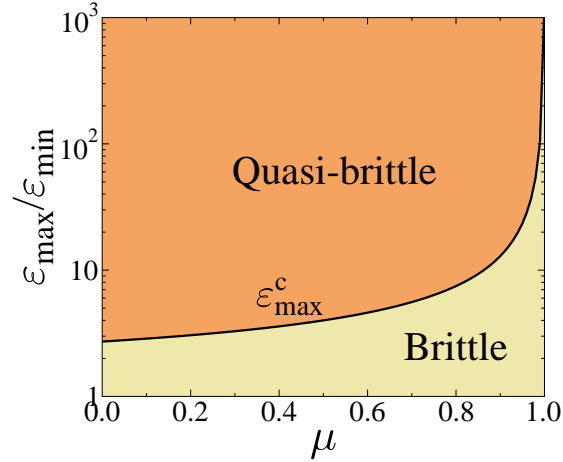}
  \caption{Phase diagram of the model on the $\mu-\varepsilon_{max}$ plane (for derivation see SI1).
  Increasing the amount of disorder the system undergoes brittle, quasi-brittle, or ductile fracture.
  The curve of $\varepsilon_{max}^c(\mu)$ gives the phase boundary between perfect brittleness and 
  quasi-brittle fracture, while ductility is obtained in the limit $\varepsilon_{max}\to+\infty$.
  (GLE 4.2.5, URL:http://glx.sourceforge.net/)
   \label{fig:phasediag}}
\end{center}
\end{figure}
The cutoff strength $\varepsilon_{max}$  takes values in the range 
$\varepsilon_{min}<\varepsilon_{max}\leq+\infty$,
while the disorder exponent $\mu$ is selected in the interval 
$0\leq\mu<1$. 
At these $\mu$ values, in the limiting case of an infinite upper cutoff 
$\varepsilon_{max}\to+\infty$ the disorder is so high in the system that no finite 
average fiber strength exists. Hence, varying the control
parameters $\varepsilon_{max}$ and $\mu$ of the model, the degree of disorder can be tuned between the extremes
of zero and infinity. 
Computer simulations were performed by slowly increasing the external load, which generates 
an intermittent dynamics where fibers break in bursts analogous to acoustic outbreaks in real experiments
(see Methods for further details of the model construction).
The size  $\Delta$ of bursts is determined as the number of fibers breaking in the correlated
trail of avalanches. 

\begin{figure}
\begin{center}
\includegraphics[bbllx=100,bblly=30,bburx=770,bbury=650,scale=0.4]{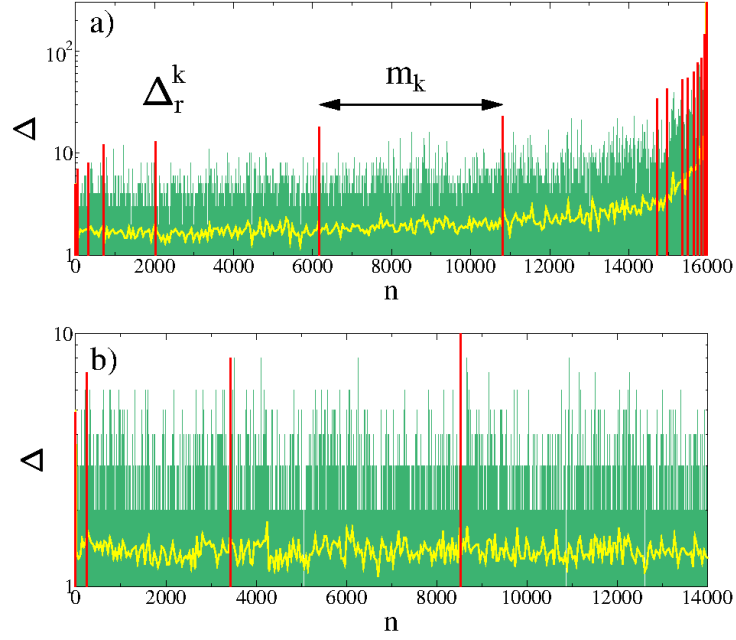}
  \caption{Event sequence of bursts, i.e.\ the size of bursts $\Delta$ is presented 
  as a function of their order number $n$ for two values of the cutoff strength 
 $(a)$ $\varepsilon_{max}=100$, and $(b)$ $\varepsilon_{max}=+\infty$
  at the disorder exponent $\mu=0.8$ in a small system of $N=10^5$ fibers. 
  The yellow line indicates the moving average of
  burst sizes $\left<\Delta\right>$ calculated over 50 consecutive events, while the red bars highlight 
  record size events of the series. The definition of the record size $\Delta_r^k$ and the 
  waiting time $m_k$ between records is also illustrated.
  (GLE 4.2.5, URL:http://glx.sourceforge.net/)
   \label{fig:sequence}}
\end{center}
\end{figure}

It is a crucial feature of our FBM that varying the amount of disorder transitions occur 
between distinct phases of perfectly brittle, quasi-brittle, and ductile behaviors with 
qualitative differences in the macroscopic response and in the precursory bursting activity. 
This enables us to quantify
how the degree of disorder affects the details of burst sequences and the forecastability of failure 
in different types of fracture.
The phase diagram in Fig.\ \ref{fig:phasediag} delineates the overall behavior of the system 
on the $\mu-\varepsilon_{max}$ plane based on analytical calculations (see Supplementary information (SI1) 
for details).
The figure demonstrates that at any value of the disorder exponent in the range $0\leq\mu<1$,
for a sufficiently low cutoff strength $\varepsilon_{max}$ of fibers
$\varepsilon_{max}<\varepsilon_{max}^c(\mu)$, the bundle exhibits a perfectly brittle response, i.e.\
a linearly elastic behavior is obtained up to 
the breaking of the weakest fiber which then triggers a catastrophic burst leading to an immediate abrupt failure.
Above the phase boundary $\varepsilon_{max}^c(\mu)$ a quasi-brittle phase emerges,
where global failure is approached through intermittent breaking avalanches. A representative example
of the sequence of bursts up to failure can be seen in Fig.\ 
\ref{fig:sequence}$(a)$, where the size of bursts $\Delta$ is presented as a function of their 
order number $n$, similarly to the natural time analysis of time series \cite{varotsos_natural_time_book,sarlis_epl_2009,sarlis_physica_2018}. 
The size of bursts $\Delta$ fluctuates due to the disorder of fibers' 
strength, however, the evolution of the overall structure of the sequence can be inferred: 
at the beginning of the fracture process the moving average of the burst size $\left<\Delta\right>$ 
remains nearly constant which shows the high degree of stationarity of the crackling activity over a broad
range. However, close to failure the rapidly increasing average burst size $\left<\Delta\right>$ and 
the growing fluctuations of $\Delta$ indicate the acceleration of the fracture process.
This generic trend of the cracking sequence is also confirmed by the behavior of the 
average number $a$ of fiber breakings
triggered immediately after the breaking of a single fiber at the strain $\varepsilon$
(for derivation see SI2)
\begin{equation}
a(\varepsilon) = \frac{\mu}{1-\left(\frac{\varepsilon}{\varepsilon_{max}}\right)^{\mu}}.
\end{equation}
The expression of $a(\varepsilon)$ has to be evaluated over the range 
$\varepsilon_{min}\leq\varepsilon\leq\varepsilon_c$,
where $\varepsilon_c$ denotes the critical strain $\varepsilon_c=\varepsilon_{max}/(1-\mu)^{1/\mu}$,
where failure occurs.
As the system approaches the critical point  $\varepsilon_c$, the value of $a$
monotonically increases to 1 indicating the acceleration of the fracture process and the instability emerging 
at the critical point.
Ultimate failure occurs in the form of a catastrophic avalanche $a\geq1$, 
where all the remaining intact fibers break
in one event.
Varying the control parameters $\varepsilon_{max}$ and $\mu$ inside the quasi-brittle phase, 
the qualitative structure of the event series remains the same, however, the extension 
of the accelerating regime and the magnitude of acceleration, which are essential features for 
forecasting, are changing. The overall behaviour of the sequence of crackling events of our model
has a nice qualitative agreement with the accelerating acoustic and seismic activity accompanying 
the creep rupture \cite{garcimar_statistical_1997,guarino_experimental_1998,nechad_creep_2005,rosti_statistics_2010,
alava_lifetime_pre2016}
and compressive failure \cite{vasseur_scirep_2015,nataf_predicting_2014,salje_prl_2018,vives_coal_burst_2019}
of various types of disordered materials, and the 
failure phenomena of geosystems such as volcanic eruptions \cite{main_GJI:GJI4982,main_limits_2013}, 
cliff collapse \cite{amitrano_seismic_2005}, breakoff of hanging glaciers \cite{faillettaz_evidence_2008}, 
and landslides \cite{sammonds_bvalue_grl1998,michlmayr_fiber_optic2017}.

Our system has the remarkable feature that in the limit of very high disorder $\varepsilon_{max}\to+\infty$,
the acceleration of the dynamics disappears $a=\mu$, and the 
entire series of crackling events remains stationary until the last avalanche. 
Figure \ref{fig:sequence}$(b)$ illustrates that the stability of the failure process is retained
until the end, and no catastrophic avalanche emerges, hence, this type of fracture process is considered to be 
ductile in the model. 
The underlying mechanism is that due to the slow decay of the fat-tailed distribution 
of fibers' strength Eq.\ (\ref{eq:fstrength}), 
there are sufficiently strong fibers 
in the bundle which can always stabilize the fracture process. Note that for $\mu>1$
the system always undergoes perfectly brittle fracture at any cutoff strength $\varepsilon_{max}$ 
(see Fig.\ \ref{fig:phasediag}).

This powerful model allows us to unveil how the changing degree of disorder 
affects the predictability of the ultimate failure of the system. 
In the limiting case of perfectly brittle fracture, the system collapses when 
its load reaches the strength of the weakest fiber. Since it is not known 
a priori, the failure point has a great uncertainty.
In the ductile phase of high disorder, 
the stationary bursting activity does not provide any hint of the imminent failure.
Tuning the amount of disorder by varying $\mu$ and 
$\varepsilon_{max}$, we can drive the system between these two limits giving a precise quantitative
characterization of the strength of acceleration of the precursory activity, 
and hence, the forecastability of failure.
Instead of the time of ultimate failure, we focus on the onset of acceleration,
analyzing the statistics of record size events of the crackling sequence.

\subsection*{Approaching failure through record breaking bursts}
\begin{figure}
\begin{center}
\includegraphics[bbllx=20,bblly=20,bburx=1100,bbury=330,scale=0.47]{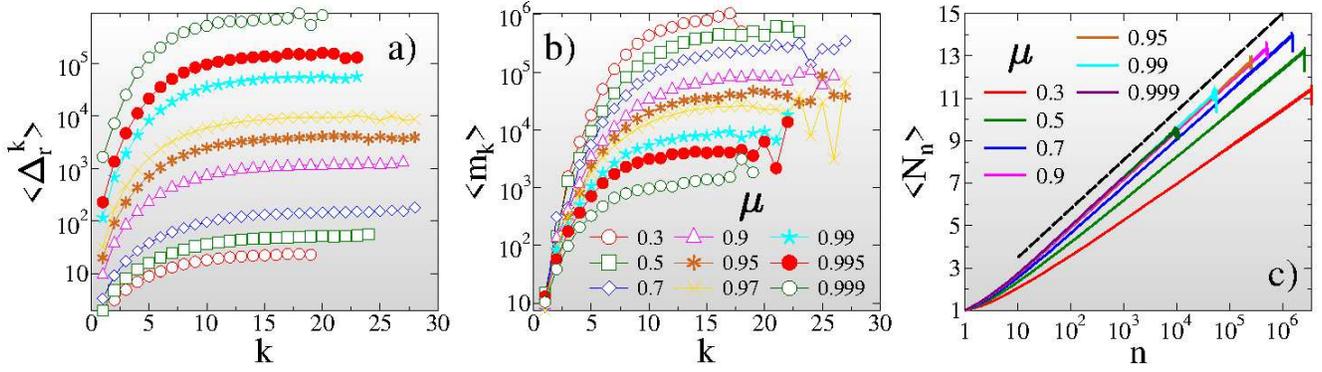}
  \caption{Average size $\left<\Delta_r^k\right>$ $(a)$ and lifetime $\left<m_k\right>$ $(b)$ of records 
  as a function of their rank $k$ for several values of the disorder exponent $\mu$ 
  at an infinite cutoff strength $\varepsilon=+\infty$. $(c)$ The average number of records $\left<N_n\right>$
  that occurred until $n$ bursts are generated during the fracture process. 
  The dashed line represents a logarithmic function.
  (GLE 4.2.5, URL:http://glx.sourceforge.net/)
   \label{fig:inf_av_recsize_wait_num}}
\end{center}
\end{figure}
A record of crackling noise is a burst which has a size $\Delta_r$ greater than 
any previous event, similarly to  
sports like athletics, where records are the best results of a discipline
\cite{records_athletics_2007}. 
Assuming that the first burst is a record, record breaking (RB) events form 
a monotonically increasing 
sub-sequence during the fracture process, as it is illustrated in Fig.\ \ref{fig:sequence}.
RB events are identified by their rank $k=1,2, \ldots$, 
which occurred as the $n_k$th event of the complete sequence with size 
$\Delta_r^k$. As fracture proceeds, records get broken after a certain number
of bursts giving rise to new records. The number of events, one has to wait
to break the $k$th record, defines the waiting time $m_k$
\begin{equation}  
m_k = n_{k+1} - n_k, \label{eq:mk} 
\end{equation} 
which can also be considered as the lifetime of the record. The definition of the record 
characteristics $\Delta_r^k$ and $m_k$ is illustrated in Fig.\ \ref{fig:sequence}$(a)$.
It can be observed that in the accelerating regime of the 
quasi-brittle fracture process in Fig.\ \ref{fig:sequence}$(a)$, 
records rapidly follow each other reaching the total number
$N_n^{tot}=22$, while in the stationary 
burst sequence of ductile failure in Fig.\ \ref{fig:sequence}$(b)$ the value of $N_n^{tot}$ 
remains significantly lower ($N_n^{tot}=4$), although the total number of events fall close to 
each other in the two cases. These small subsets of RB events grasp key aspects of the dynamics 
of fracture, namely, the acceleration of fracturing towards failure is accompanied by an
accelerated record breaking indicated by the decreasing waiting time $m_k$ in Fig.\ 
\ref{fig:sequence}$(a)$. However, 
in the stationary regime of the beginning of the fracture process, record breaking slows down 
with increasing values of $m_k$, similarly to ductile fracture in Fig.\ \ref{fig:sequence}$(b)$.
To quantify these important trends and correlations in burst sequences,
we analyze the statistics of the size and lifetime of records, and their evolution with the record rank
as the system approaches failure at different degrees of disorder. 

\subsubsection*{Record statistics in the limit of high disorder}
Figure \ref{fig:inf_av_recsize_wait_num} demonstrates that in ductile fracture $\varepsilon_{max}=+\infty$
the statistics of records is entirely consistent with the behavior
of sequences of independent identically distributed (IID) random variables
\cite{wergen_record_2013,scherbakov_pre_2013}: 
The average record size $\left<\Delta_r^k\right>$ is of course a monotonically increasing 
function of the record rank $k=1, 2, \ldots $ for all values of the disorder exponent $\mu$ 
(Fig.\ \ref{fig:inf_av_recsize_wait_num}$(a)$). 
\begin{figure}
\begin{center}
\includegraphics[bbllx=20,bblly=20,bburx=740,bbury=650,scale=0.4]{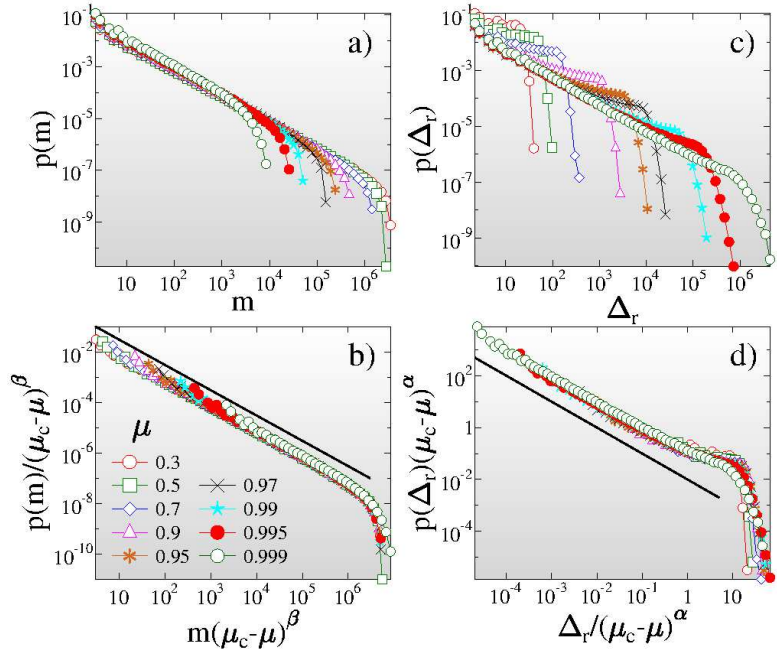}
  \caption{Probability distribution of the lifetime $p(m)$ $(a)$ and size $p(\Delta_r)$ $(c)$ 
  of records for several values of the disorder exponent $\mu$ at $\varepsilon=+\infty$. Rescaling the 
  distributions with proper powers $\alpha$ and $\beta$ of the distance from the critical point $1-\mu$,
  distributions obtained at different $\mu$ values can be collapsed on a master curve. In $(b)$ and 
  $(d)$ the straight lines represent power laws of exponent $z=1$ and $\tau_r=1$, respectively. 
  The legend is presented in $(b)$   for all the figures.
  (GLE 4.2.5, URL:http://glx.sourceforge.net/)
   \label{fig:inf_dists}}
\end{center}
\end{figure}
The average lifetime of records $\left<m_k\right>$ also increases with $k$, 
since it gets more and more difficult to break 
the growing records of a stationary sequence (Fig.\ \ref{fig:inf_av_recsize_wait_num}$(b)$). 
It follows that as fracture proceeds, the number of records $N_n$
slowly increases with the total number of bursts $n$.
For IID sequences a universal logarithmic dependence has been derived 
for the average number of records $\left<N_n\right>$ that occurred until the $n$th event of the sequence
\cite{wergen_record_2013}.
Figure \ref{fig:inf_av_recsize_wait_num}$(c)$ shows that the IID result perfectly describes the record 
breaking process of ductile fracture 
\begin{equation}
\left<N_n\right> \approx A+B\ln n, 
\label{eq:log_depend}
\end{equation}
with the additional feature that the multiplication factor $B$ of the logarithmic term depends on $\mu$.
Reducing the amount of disorder in the ductile phase by increasing the exponent $\mu$ 
towards the critical point of perfectly brittle failure $\mu\to\mu_c(\varepsilon_{max}^c=+\infty)=1$, 
the qualitative behavior of the curves in Fig.\ \ref{fig:inf_av_recsize_wait_num} 
remains the same, however, record breaking accelerates:
the asymptotic value of the record size in Fig.\ \ref{fig:inf_av_recsize_wait_num}$(a)$
rapidly increases, while the asymptotic record lifetime in Fig.\ 
\ref{fig:inf_av_recsize_wait_num}$(b)$ tends to zero as $\mu$ approaches 1.
We quantified this behavior by calculating the average value of the largest record size 
$\left<\Delta_r^{max}\right>$ and largest waiting time $\left<m^{max}\right>$ that occurred 
up to failure as function of $\mu$. Both quantities proved to have a power law dependence on 
the distance from the critical point
\begin{eqnarray}
\left<\Delta_r^{max}\right> \sim (\mu_c-\mu)^{-\alpha}, \qquad \qquad \left<m^{max}\right> \sim (\mu_c-\mu)^{\beta},
\label{eq:alpha_beta}
\end{eqnarray}
with the critical exponents $\alpha=1.8\pm 0.05$ and $\beta=1.0\pm 0.05$ (for figure see SI3). 
The acceleration of record breaking has also the consequence that the number of records 
$\left<N_n\right>$ grows faster with the event number $n$ for higher $\mu$, 
i.e.\ the multiplication factor $B$ in Eq.\ (\ref{eq:log_depend}) increases to 1 as $\mu$ 
approaches $\mu_c=1$ but the IID nature of the event series is retained.

The IID behavior is also confirmed by the overall statistics of the lifetime $m$ 
of records. The probability distribution $p(m)$ of the record lifetime $m$ has
a power law functional form  
\begin{equation}
p(m)\sim m^{-z},
\label{eq:pmiid}
\end{equation}
with an exponent $z$, which has a universal value $z=1$ equal to its IID counterpart \cite{wergen_record_2013,scherbakov_pre_2013} (see Fig.\ \ref{fig:inf_dists}$(a)$).
For the size distribution of records $p(\Delta_r)$ a non-universal behavior is expected in the sense that 
$p(\Delta_r)$ depends on the underlying distribution of burst sizes \cite{wergen_record_2013,scherbakov_pre_2013}. 
In our system a power law distribution is obtained 
\begin{equation}
p(\Delta_r)\sim\Delta_r^{-\tau_r},
\label{eq:recsizedist_fin}
\end{equation}
with an exponent $\tau_r=1$, which does not depend on the value of $\mu$ (see Fig.\ \ref{fig:inf_dists}$(c)$). 
It is important to emphasize that approaching the critical point of brittle failure the 
cutoff of the size distributions $p(\Delta_r)$ diverges, while the one of the lifetime distribution $p(m)$
tends to zero, in agreement with the behavior of the high rank limit of the average size and 
lifetime of records in Figs.\ \ref{fig:inf_av_recsize_wait_num}$(a,b)$.
Figures \ref{fig:inf_dists}$(b,d)$ demonstrate that rescaling the distributions $p(\Delta_r)$ and 
$p(m)$ according to the scaling laws Eq.\ (\ref{eq:alpha_beta}), the curves obtained at different
$\mu$ exponents can be collapsed on the top of each other. The good quality data collapse 
confirms the consistency of the results.
\begin{figure}
\begin{center}
\includegraphics[bbllx=30,bblly=20,bburx=730,bbury=330,scale=0.45]{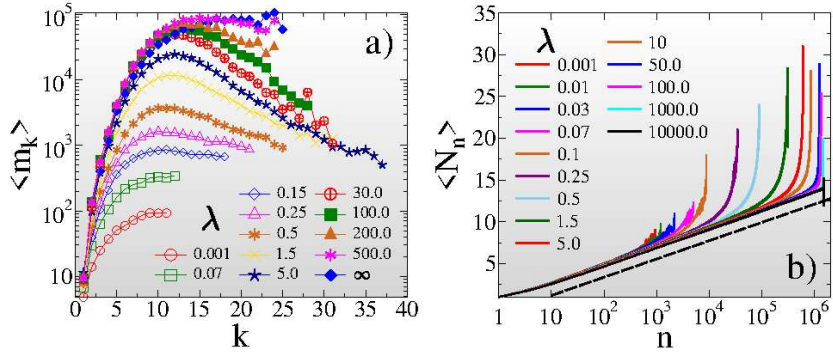}
  \caption{$(a)$ Average lifetime of records $\left<m_k\right>$ as a function of the record rank $k$
  for several values of the cutoff strength $\lambda$. $(b)$ Average number of records $\left<N_n\right>$
  formed during the fracture process as a function of the event number $n$ varying $\lambda$
  in a broad range. The value of the disorder exponent is fixed to $\mu=0.7$.
  (GLE 4.2.5, URL:http://glx.sourceforge.net/)
   \label{fig:fin_averwait_recnum}}
\end{center}
\end{figure}
The excellent agreement of the record statistics of the burst sequence of ductile fracture 
with the behavior of IID sequences implies that, in spite of the increasing external load 
on the bundle,
the entire fracture process is controlled by the microscale disorder
of the system. 

\subsubsection*{Approaching failure through accelerated record breaking}

Inside the quasi-brittle phase, when the amount of disorder is reduced by the finite cutoff 
strength $\varepsilon_{max}$ of fibers, the evolution of burst sequences substantially changes since the
initial stationary regime is followed by acceleration in the vicinity of failure (see
Fig.\ \ref{fig:sequence}$(a)$). To facilitate the comparison of results obtained varying the 
cutoff strength $\varepsilon_{max}$ at different exponents $\mu$,
we introduce the parameter $\lambda=(\varepsilon_{max}-\varepsilon_{max}^c)/\varepsilon_{max}^c$, 
which characterizes the relative distance $\lambda>0$ of the system 
from the phase boundary $\varepsilon_{max}^c(\mu)$ at a given value of $\mu$. 
In order to determine how the precise amount of disorder controls the onset and significance 
of acceleration, and hence, the forecastability of ultimate failure, we performed computer simulations
varying the value of $\lambda$ and the disorder exponent $\mu$ in broad ranges 
$0.001\leq\lambda\leq +\infty$ and $0.01\leq\mu\leq 1$, respectively. At each parameter 
set averages were calculated over 6000 samples (for details of averaging see Methods).

Of course, the average size of records $\left<\Delta_r^k\right>$ 
has the same monotonically increasing trend with the record rank $k$ as in the ductile 
phase for all values of $\lambda$ and $\mu$ (see SI4 for figure). 
However, the average lifetime of records $\left<m_k\right>$ exhibits an astonishing behavior: 
As a representative example, Fig.\ \ref{fig:fin_averwait_recnum}$(a)$ 
demonstrates for $\mu=0.7$ that for a sufficiently high 
disorder $\lambda> 0.1$ 
the $\left<m_k\right>$ curves develop a maximum at a characteristic record rank $k^*$. 
For low rank records $k<k^*$ the RB process slows 
down due to the stationarity of the beginning of the fracture process, while 
beyond the maximum  $k>k^*$, the decreasing lifetime 
indicates that the approach to failure is accompanied by an accelerated record breaking.
Note that as the amount of disorder grows with increasing 
$\lambda$, the position of the maximum $k^*$ slightly shifts to higher values,
while the maximum gets gradually less pronounced and eventually disappears when approaching
ductile fracture. 
\begin{figure}
\begin{center}
\includegraphics[bbllx=30,bblly=15,bburx=1100,bbury=330,scale=0.47]{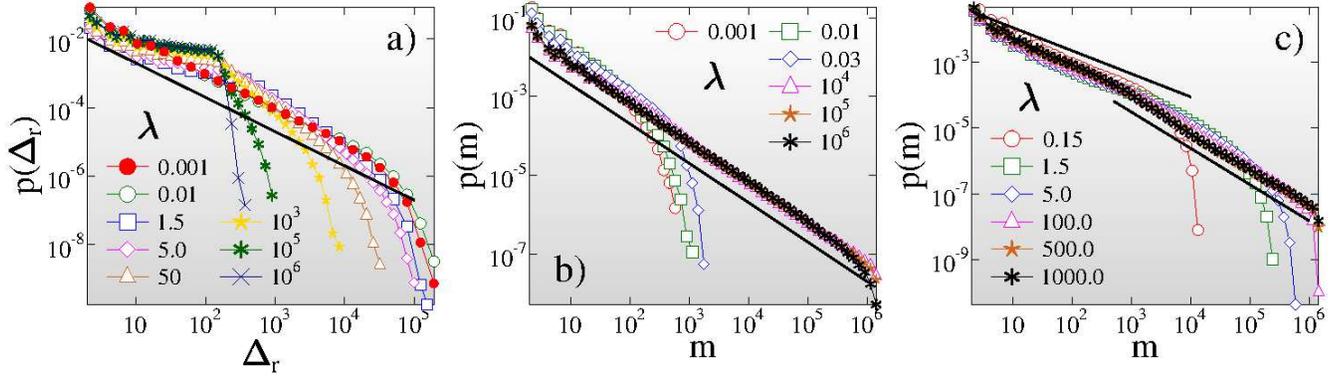}
  \caption{$(a)$ Size distribution of record events $p(\Delta_r)$ for several values 
  of $\lambda$. The straight line represents a power law of exponent $\tau_r=1$.
  $(b)$ Probability distribution of the lifetime of records $p(m)$ for  
  cutoff strengths of fibers $\lambda$ where the average waiting time in Fig.\ 
  \ref{fig:fin_averwait_recnum}$(b)$ solely exhibits slowdown. The straight line
  represents a power law of exponent $z=1$.
  $(c)$ Lifetime distributions $p(m)$ for intermediate $\lambda$ values where the RB process accelerates 
  prior to failure. The two straight lines represent power laws of exponent $z_{a1}=0.7$ and $z_{a2}=1.15$.
  The disorder exponent $\mu$ has the same value $\mu=0.7$ in the figures. 
  (GLE 4.2.5, URL:http://glx.sourceforge.net/)
   \label{fig:fin_dist_combine}}
\end{center}
\end{figure}
Surprisingly, in the opposite limit $\lambda\to 0$ of low disorder,  
a similar behavior is observed: the record lifetime $\left<m_k\right>$ 
tends to a monotonically increasing form showing the slowdown of record breaking, in spite of the 
rapid collapse of the highly brittle system.

The acceleration of the RB process results in a rapid increase of the number $N_n$
of records close to failure. Figure \ref{fig:fin_averwait_recnum}$(b)$ shows that at early 
stages of the fracture process the average record number $\left<N_n\right>$
increases logarithmically with the number of bursts $n$ in agreement with Eq.\ (\ref{eq:log_depend}), 
however, at a characteristic 
event index $n_{k^*}$, corresponding approximately to the rank $k^*$ of maximum lifetime, 
the functional form of $\left<N_n\right>$ changes to a rapid increase.
It is important to emphasize that varying the amount of disorder by $\lambda$,
as the accelerating regime diminishes in the limits of highly brittle ($\lambda\to 0$) 
and ductile ($\lambda\to+\infty$) fracture, the deviations 
from the generic logarithmic trend of IIDs gradually disappear apart from some fluctuations.
 
Also the overall statistics of record sizes and lifetimes
is sensitive to the degree of disorder. 
Figure \ref{fig:fin_dist_combine}$(a)$ shows
that close to the phase boundary of perfect brittleness $\lambda\to 0$ the size distribution of records
$p(\Delta_r)$ is identical with the corresponding distribution of ductile fracture  
(see Fig.\ \ref{fig:inf_dists}$(c)$ for comparison), 
i.e.\ a power law distribution Eq.\ (\ref{eq:recsizedist_fin}) 
is obtained with a universal exponent $\tau_r=1$.
Further from the phase boundary, the change in the dynamics of fracture from a steady evolution to 
acceleration gives rise to a crossover of $p(\Delta_r)$ between two 
regimes of different functional forms: for small records, generated during early stages of 
the fracture process, the distribution remains similar to its ductile counterpart, 
however, beyond a characteristic record size a steeper power law is formed.
Simulations revealed that the crossover point
practically coincides with the average record size $\left<\Delta_r^k\right>$ at the rank 
$k^*$ of the maximum lifetime in Fig.\ \ref{fig:fin_averwait_recnum}$(a)$.  
The underlying mechanism of the emergence of the crossover of the distribution of record sizes 
is the crossover of the entire burst size distribution \cite{kadar_kun_pre2019},
similar to the so-called b-value anomaly observed for event magnitude distributions in 
geosystems \cite{sammonds_role_1992,amitrano_2012_epjst,amitrano_brittle-ductile_2003}.

The lifetime distribution $p(m)$ shows the same qualitative behavior as $\lambda$ is varied. 
For clarity, the distributions $p(m)$ are 
presented separately for the parameter ranges where the acceleration of record breaking is absent 
and present, in Fig.\ \ref{fig:fin_dist_combine}$(b)$ and $(c)$, respectively.
At low and high disorder, where no acceleration occurs,
the waiting time distributions are consistent with the IID behavior Eq.\ (\ref{eq:pmiid}) (Fig.\ \ref{fig:fin_dist_combine}$(b)$).
For intermediate $\lambda$, when accelerated record breaking emerges, $p(m)$ exhibits 
a crossover between two power laws of different exponents (Fig.\ \ref{fig:fin_dist_combine}$(c)$).
For short lifetimes, typical for the vicinity of failure, the
exponent $z_{a1}$ is smaller $z_{a1}=0.7$, while, for large waiting times characteristic for the 
initial slowdown of the process, the exponent is higher  $z_{a2}=1.15$ than the IID result 
$z=1$. It can be observed in Figs.\ \ref{fig:fin_dist_combine}$(b,c)$ that the particular 
value of $\lambda$ inside the two regimes 
only affects the crossover point and the cutoff of the distributions.

Simulations revealed that varying the disorder exponent $\mu$ the qualitative behaviour 
of the average waiting time $\left<m_k\right>$ and event number $\left<N_n\right>$,
and of the distributions of record sizes $p(\Delta_r)$ and lifetimes $p(m)$, remains the same,
they undergo only quantitative changes what we analyze in the next section.
The results demonstrate that the statistics of records is very sensitive to the details of the 
crackling sequence providing a powerful tool to identify the onset of acceleration towards failure. 
Additionally, comparing the results
to IID event sequences, our analysis proves that 
before acceleration the precursory bursting activity is dominated
by the microscale disorder giving rise to stationarity. 
Acceleration starts when the stress enhancements generated by 
the redistribution of load after breaking bursts, can enhance the triggering 
of further avalanches.

\section*{The effect of disorder on the onset of acceleration}

The presence of a sufficiently broad accelerating regime in the series of crackling events
is of ultimate importance to obtain an early warning of the imminent failure and to forecast the 
final collapse of the evolving system with a good precision. The position of the maximum $k^*$ of the average 
record lifetime $\left<m_k\right>$ and the corresponding event index $n_{k^{*}}$ 
provide an excellent signal of the start of acceleration 
towards failure after a longer period of stationary evolution.
However, it can be observed in Fig.\ 
\ref{fig:fin_averwait_recnum}$(a)$ that both the extension of the accelerating regime and the
magnitude of acceleration depend on the degree of disorder in the system.
To assess the predictability of the ultimate failure, the significance 
of the acceleration regime has to be characterized.

\begin{figure}
\begin{center}
\includegraphics[bbllx=10,bblly=80,bburx=390,bbury=640,scale=0.48]{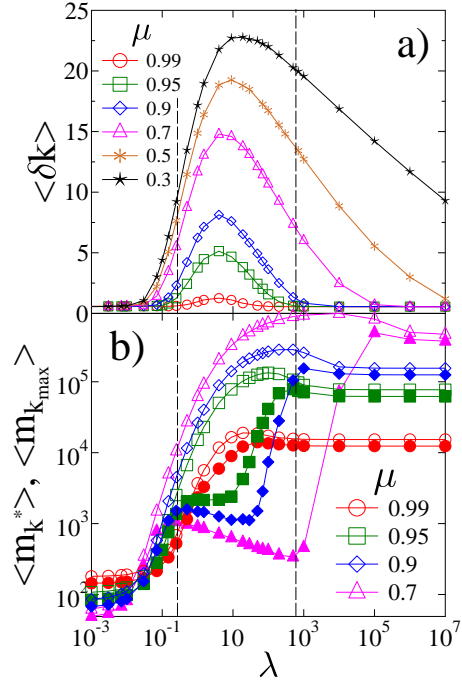}
  \caption{$(a)$ Average value of the difference $\left<\delta k\right>$ of the highest 
  record rank $k_{max}$ and the position 
  of the maximum $k^*$ of the record lifetime as a function of 
  $\lambda$ for several values of the disorder exponent $\mu$. $(b)$ The average of the maximum lifetime
  $\left<m_{k^*}\right>$ (open symbols) and the lifetime of the last record $\left<m_{k_{max}}\right>$ 
  (filled symbols) as a function of $\lambda$. For clarity, pair of curves are presented only for 
  four values of the disorder exponent $\mu$. The vertical dashed lines indicate the $\lambda$ 
  window of acceleration for $\mu=0.9$. 
  (GLE 4.2.5, URL:http://glx.sourceforge.net/)
 \label{fig:maxrank_maxpos_lambda}}
\end{center}
\end{figure}

To quantify the extension of the accelerating regime we determined the difference $\delta k$ of the 
highest record rank $k_{max}$ obtained up to failure and the position of the maximum $k^*$ of the 
record lifetimes in single samples. The average of this quantity
\begin{equation}
\left<\delta k\right>=\left<k_{max}-k^*\right>
\end{equation}
over a large number of simulations is presented in Fig.\ \ref{fig:maxrank_maxpos_lambda}$(a)$
as a function of $\lambda$ for several values of the disorder exponent $\mu$.
For each value of $\mu$ a well defined range of the cutoff strength $\lambda$
can be identified where a significant difference $\left<\delta k\right> > 1$ emerges between 
$k^*$ and $k_{max}$.
Both for a high degree of brittleness $\lambda\to0$ and for ductile breaking $\lambda\to+\infty$, 
the value of $\left<\delta k\right>$ is practically zero, which shows 
that no acceleration can be detected. 
Increasing the amount of disorder by decreasing the exponent $\mu$, 
the acceleration regime $\left<\delta k\right> > 1$ extends to higher values of the cutoff strength 
$\lambda$.

The magnitude of acceleration of the RB process can be characterized 
by comparing the last record lifetime $m_{k_{max}}$ and the maximum lifetime of records 
$m_{k^*}$. Figure \ref{fig:maxrank_maxpos_lambda}$(b)$ presents the average values 
$\left<m_{k_{max}}\right>$ and $\left<m_{k^*}\right>$, 
using the same scale of $\lambda$ on the horizontal axis
as in Fig.\ \ref{fig:maxrank_maxpos_lambda}$(a)$ to facilitate the comparison with the behavior
of $\left<\delta k\right>$. 
In the vicinity of $\mu=1$ the two quantities $\left<m_{k_{max}}\right>$ and $\left<m_{k^*}\right>$ 
practically coincide for all values 
of the cutoff strength $\lambda$. However, lowering the exponent $\mu$, 
a broader and broader range of $\lambda$ emerges where a significant difference 
 $\left<m_{k^*}\right>\gg\left<m_{k_{max}}\right>$ is obtained,
indicating the presence of a dominating maximum of the record lifetime. 
The $\lambda$ window of $\left<\delta k\right> > 1$ coincides with that of 
$\left<m_{k^*}\right>\gg\left<m_{k_{max}}\right>$ verifying the existence of a well
defined range of disorder where failure can be foreseen.
Outside this window the degree of disorder is either too low or too high so that no signal 
of the imminent failure can be identified. This $\lambda$ window of significant acceleration 
is highlighted by the vertical dashed lines in Figs.\ \ref{fig:maxrank_maxpos_lambda}$(a,b)$
for $\mu=0.9$.

\begin{figure}
\begin{center}
\includegraphics[bbllx=40,bblly=30,bburx=730,bbury=330,scale=0.43]{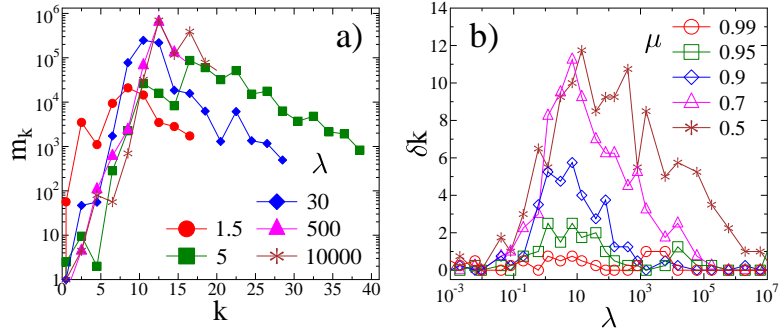}
  \caption{$(a)$ The lifetime of consecutive records $m_k$, obtained during the evolution of a 
  single system, as a function of their rank $k$ for 
  several values of $\lambda$ at the same disorder exponent $\mu=0.7$ as in Fig.\ \ref{fig:fin_averwait_recnum}$(a)$ for the sample averaged curves.
  $(b)$ The difference $\delta k$ of the highest record rank $k_{max}$ and the position 
  of the maximum $k^*$ of the record lifetime for a single sample 
  as a function of $\lambda$ for several 
  values of the disorder exponent $\mu$.
  (GLE 4.2.5, URL:http://glx.sourceforge.net/)
 \label{fig:single_sample}}
\end{center}
\end{figure}

For the practical applications of record statistics to determine the start of the 
critical regime of the dynamics, the method has to provide reliable results for single samples. Figure
\ref{fig:single_sample}$(a)$ presents the lifetime of consecutive records $m_k$ as a function
of their rank $k$ obtained by the analysis of a single system for several values of $\lambda$
at the same disorder exponent $\mu=0.7$ as in Fig.\ 
\ref{fig:fin_averwait_recnum}$(a)$. It is important to emphasize that apart from fluctuations 
records of the single sample exhibit the same overall behaviour as the sample averaged 
curves in Fig.\ \ref{fig:fin_averwait_recnum}$(a)$, i.e.\ $m_k$ has a well
defined maximum in the quasi brittle regime of the fracture process so that the value of 
the record rank $k^*$ and the corresponding event index $n_{k^*}$ can be obtained in a reliable 
way. This result is further supported by the behaviour of $\delta k$ in Fig.\ 
\ref{fig:single_sample}$(b)$ which shows that the window of forecastability of a single system,
where a significant accelerating regime can be detected by the method of record statistics,
agrees very well with the sample averaged result of Fig.\ \ref{fig:maxrank_maxpos_lambda}$(a)$.
Simulations showed that the relative value of sample-to-sample fluctuations of 
record quantities vary between 0.05 and 0.2, typically increasing with the record rank $k$,
making the method efficient for single samples undergoing quasi-brittle fracture.

\section*{Discussion}

Catastrophic failure of engineering constructions and of geosystems is often caused by the fracture 
of disordered materials. Under a constant or slowly varying external load, the inherent disorder 
of materials gives rise to a jerky evolution of the fracture process accompanied by a sequence of
acoustic outbreaks. Methods of failure forecasting predict the lifetime of the evolving system
by exploiting the power law acceleration of the precursory crackling activity prior to ultimate failure. 
Here we focused on the acceleration preceding the final collapse, and proposed a method to identify
the onset of this critical regime of the dynamics which can be used as an early warning 
of the imminent failure. 
Based on a fiber bundle model of fracture phenomena, we studied
the statistics of record size events to reveal how the sequence of breaking bursts evolves
as the system approaches failure. We demonstrated that the small subset 
of record bursts grasps essential features of the dynamics leading to ultimate failure, with the 
additional advantage that they are relatively easy to identify even experimentally over the 
noisy background. 

As an important outcome of the work, we showed that during quasi-brittle fracture, where failure
proceeds through an intense precursory activity, the acceleration of the 
crackling sequence towards failure is accompanied 
by an accelerated record breaking. The onset of acceleration 
can be identified by the record which has the longest lifetime so that its rank $k^*$, and the corresponding 
event index $n_{k^{*}}$ provide a reliable signal in the burst sequence where the critical regime
of the dynamics starts. Before this characteristic record, the process of record breaking slows down
and the statistics of records proved to be equivalent with the behavior of event sequences of IIDs.
These results imply that the beginning of fracture is dominated by the disorder of the 
material in spite of the increasing external load and decreasing load bearing capacity of the system. 
Beyond $k^*$, acceleration is caused by the enhanced triggering of bursts following 
the stress redistribution after breaking events.

Early warning and forecastability of the imminent failure requires a sufficiently
broad critical regime with a considerable magnitude of acceleration. To asses the 
effect of disorder on the forecastability of failure,
we made a quantitative characterization of the significance of acceleration in terms of 
record statistics. Most notably, we showed that the highly brittle fracture of low disorder materials, and 
the ductile failure of the strongly disordered ones, are both unpredictable. In spite 
of the considerable number of bursts generated, the absence of acceleration limits forecastability to  a well
defined range of disorder on the phase diagram of the system. 
\begin{figure}
\begin{center}
\includegraphics[bbllx=50,bblly=20,bburx=600,bbury=450,scale=0.46]{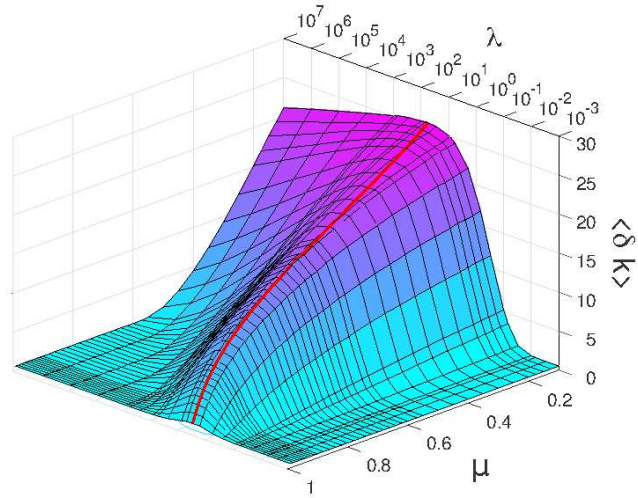}
  \caption{Three-dimensional representation of the 
  $\left<\delta k(\mu,\lambda)\right>$ function, which gives an overview of the forecastability 
  of failure in terms of disorder. The bold red line highlights the ridge of the surface, where the 
  broadest acceleration regime is obtained.
  (MATLAB 2018a, URL:https://uk.mathworks.com/)
 \label{fig:3dsurface}}
\end{center}
\end{figure}
This is illustrated in Fig.\ 
\ref{fig:3dsurface} which presents the value of $\left<\delta k\right>$ over the $\mu-\lambda$
plane. It can be observed that for the significance of the accelerating regime,
disorder has an optimum amount, i.e.\ the combination of the disorder exponent $\mu$ 
and of the cutoff strength $\lambda$ of fibers determining the ridge of the $\left<\delta k\right>$ surface
provides the best predictability. Note that lowering the exponent $\mu$ results in a broadening 
of the $\lambda$ range where a significant acceleration occurs so that the range of predictability 
tends to infinity in terms of $\lambda$ for $\mu\to0$. 
Our results imply that former conclusions in the literature that increasing disorder improves forecastability
is generally not valid for fat-tailed disorder. Increasing the cutoff strength $\lambda$ beyond 
the ridge of $\left<\delta k\right>(\mu,\lambda)$ at fixed exponents $\mu$ in Fig.\
\ref{fig:3dsurface}, disorder becomes disadvantageous. The absolute upper bound
of predictability is defined by the line on the $\mu-\lambda$ plane, where $\left<\delta k\right>\approx 0$
holds in the high $\lambda$ regime (see Fig.\ \ref{fig:3dsurface}). This bound
emerges due to the slow decay of the fat-tailed threshold distribution: since the tail of the 
strength distribution Eq.\ (\ref{eq:fstrength}) is efficiently sampled even at 
small system sizes, at sufficiently high upper cutoffs
$\lambda$ there will be so strong fibers in the system which can stabilize the fracture process till
the end. As a consequence, ductile behaviour can already be reached at finite $\lambda$ values.
Predictability has also a lower bound which extends from $\lambda\approx 0.01$ to $\lambda\approx 0.1$
as $\mu$ increases from 0 to 1 (see Fig.\ \ref{fig:3dsurface}).

To analyze the evolution of the failure process, we only considered the magnitude of crackling events.
The RB analysis is similar to the natural time analysis in the sense that the physical time of events 
is ignored and only the magnitude of events is considered as a function of their order parameter.
For practical applications, the adaptation of our method is straightforward in those cases,
where the system approaches failure through an increasing activity of acoustic or seismic events. 
Laboratory experiments have revealed an accelerating rate of acoustic events 
accompanied by an increasing average event magnitude during the tertiary regime 
of the creep failure of heterogeneous materials
\cite{garcimar_statistical_1997,guarino_experimental_1998,nechad_creep_2005,rosti_statistics_2010,
alava_lifetime_pre2016}, and for the approach to failure 
of porous rocks under a slowly increasing compressive load
\cite{vasseur_scirep_2015,nataf_predicting_2014,salje_prl_2018,vives_coal_burst_2019}. 
A similar behaviour of the rate of acoustic or seismic signals 
has been observed for the failure of geosystems such as volcanic eruptions
\cite{main_GJI:GJI4982,main_limits_2013}, 
cliff collapses \cite{amitrano_seismic_2005}, the breakoff of hanging glaciers
\cite{faillettaz_evidence_2008}, and in some cases also for landslides 
\cite{sammonds_bvalue_grl1998,michlmayr_fiber_optic2017}. In these systems, records either of the 
fluctuating daily (hourly) rate of seismic (acoustic) events, or of the magnitude of individual signals 
can serve as the starting point of the analysis of record statistics. 
After setting the first record of the time window of the analysis,
the rest of the record events can be unambigously identified since the 
largest events of the sequence have to be found. Our method suggests that the event index $n_{k^*}$
of the longest living record, and its corresponding physical time, provide the onset of 
acceleration of the record breaking process, and hence, of the start of the critical regime of
the approach to failure.
Failure forecast methods (FFM) predict the time of failure based on the power law acceleration of a 
characteristic quantity of the time evolution of the system. Our method can also be used to complement FFMs
by conditioning a series of discrete events to identify the time window where the assumption of a power 
law acceleration is applicable \cite{voight_method_1988,main_ffm_1999}.

Our study is based on a fiber bundle model with equal load sharing which is essentially a 
mean field approach to fracture. The model has the advantage that solely one source of 
disorder is present in the system, i.e.\ the random strength of fibers. In more realistic situations
stress fluctuations occur around cracks. Recently, we have shown in Ref.\ \cite{danku_disorder_2016}
that when the strength disorder is very high in the system with fat-tailed local strength distributions,
stress concentration around cracks have a minor effect on the breakdown process. Hence, we
conjecture that our statements have a broader validity, they are not limited by the assumption of 
the homogeneous stress field.

\section*{Methods}
To study the fracture of heterogeneous materials we use a fiber bundle model which provides 
a straightforward way to reveal the role of microscale disorder in fracturing. The model is composed 
of $N$ parallel fibers with a perfectly brittle response, i.e.\ the fibers exhibit a 
linearly elastic behavior up to breaking at a critical deformation $\varepsilon_{th}$. 
Materials' disorder is introduced by the random strength of fibers 
$\varepsilon_{th}^i (i=1,\ldots , N)$ for which we considered a power law distribution
over a finite range. Computer simulations were performed by slowly increasing the external load 
on the bundle to provoke the breaking of a single fiber. 
The load of broken fibers is overtaken by the remaining intact ones. 
We assume equal load sharing after breaking events which ensures that all fibers 
keep the same load during the entire breaking process. Since no stress fluctuations can arise, 
the random strength of fibers is the only source of disorder in the system.

After load redistribution, the excess load may trigger additional breakings, which is again
followed by load redistribution. Eventually, such repeated cycles of breaking and load 
redistribution steps give rise to bursts of breakings in the model which are analogous 
to the acoustic outbreaks of real experiments. The size of bursts $\Delta$ is defined as the 
number of fibers breaking in the avalanche. 
In all the simulations presented in the manuscript the number of fibers was fixed $N=5\times 10^6$
and averaging was performed over 6000 samples, which provided a sufficient precision for the data
analysis. 

Characteristic quantities of records such as the average size 
$\left<\Delta^k_r\right>$ and lifetime $\left<m_k\right>$ were obtained by averaging over the samples
at fixed record ranks $k$. The same type of sample average was calculated for the average 
number of records $\left<N_n\right>$ and $\left<\delta k\right>$ at fixed event numbers $n$
and upper cutoffs $\lambda$, respectively. Probability densities $p(\Delta_r)$ and $p(m)$ 
were obtained for the entire ensemble of samples at given values of the control parameters $\mu$
and $\lambda$.

\bibliography{/home/feri/papers/statphys_fracture}

\section*{Acknowledgments}
The work is supported by the EFOP-3.6.1-16-2016-00022 project. 
The project is co-financed by the European Union and the European Social Fund.
This research was supported by the National Research, Development and
Innovation Fund of Hungary, financed under the K-16 funding scheme 
Project no.\ K 119967.
The research was financed by the Higher Education Institutional
Excellence Program of the Ministry of Human Capacities in Hungary, 
within the framework of the Energetics thematic
program of the University of Debrecen.

\section*{Author contributions statement}
VK carried out computer simulations. VK, GP, and FK performed the data analysis. 
FK conceived of and designed the study, and drafted the manuscript. 
All authors read and approved the manuscript.
\section*{Additional information}

\textbf{Competing interests} 
The authors declare no competing financial interests.

The corresponding author is responsible for submitting a \href{http://www.nature.com/srep/policies/index.html#competing}{competing interests statement} on behalf of all authors of the paper.

\end{document}